\documentclass[runningheads]{llncs}
\usepackage[T1]{fontenc}
\usepackage{graphicx}
\usepackage{tikz}
\usepackage[utf8x]{inputenc}
\usepackage[english]{babel}
\usepackage{enumitem}
\usepackage{booktabs} 
\usepackage{tcolorbox}
\usepackage{tabularx}
\usepackage{hyperref}

\usepackage{academicons}
\definecolor{orcidlogocol}{HTML}{A6CE39}
\newcommand{\orcid}[1]{\href{https://orcid.org/#1}{\includegraphics{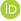}}}

\newif\ifanonymous
\anonymousfalse

\begin{document}
\title{Time-Based State-Management of Hash-Based Signature CAs for VPN-Authentication}
\titlerunning{Time-Based State-Management for VPN-Authentication}
%
\ifanonymous
\author{Anonymous}
\authorrunning{Anonymous}
\institute{Anonymous Institution}
\else
\author{Daniel Herzinger\inst{1}\orcid{0009-0009-3775-9567} \and
Linus Heise\inst{2}\orcid{0009-0004-0017-5828} \and Daniel
Loebenberger\inst{2}\orcid{0000-0002-7969-6260} \and Matthias Söllner\inst{2}}
\authorrunning{D. Herzinger et al.}
%
\institute{genua GmbH, Germany\\ \email{daniel\_herzinger@genua.de} \and
	OTH Amberg-Weiden, Germany\\ \email{\{l.heise, d.loebenberger, m.soellner\}@oth-aw.de}}
\fi
\maketitle              

\begin{abstract}
Advances in quantum computing necessitate migrating the entire technology stack to post-quantum cryptography. This includes IPsec-based VPN connection authentication. Although there is an RFC draft for post-quantum authentication in this setting, the draft does not consider (stateful) hash-based signatures despite their small signature size and trusted long-term security.
\newline
We propose a design with time-based state-management that assigns VPN devices a
certificate authority~(CA) based on the hash-based signature scheme XMSS\@. The CA then issues leaf certificates which are based on classical cryptography but have a short validity time, e.\,g., four hours. It is to be expected that even large quantum computers will take significantly longer to break the cryptography, making the design quantum-secure. We propose strategies to make the timekeeping more resilient to faults and tampering, as well as strategies to recognize a wrong system time, minimize its potential damage, and quickly recover.
\newline
The result is an OpenBSD implementation of a quantum-safe and, regarding the leaf certificates, highly flexible VPN authentication design that requires significantly less bandwidth and computational resources compared to existing alternatives.

\end{abstract}
\section{Introduction}
\begin{figure}[t]
	\center
	\includegraphics[width=0.9\textwidth]{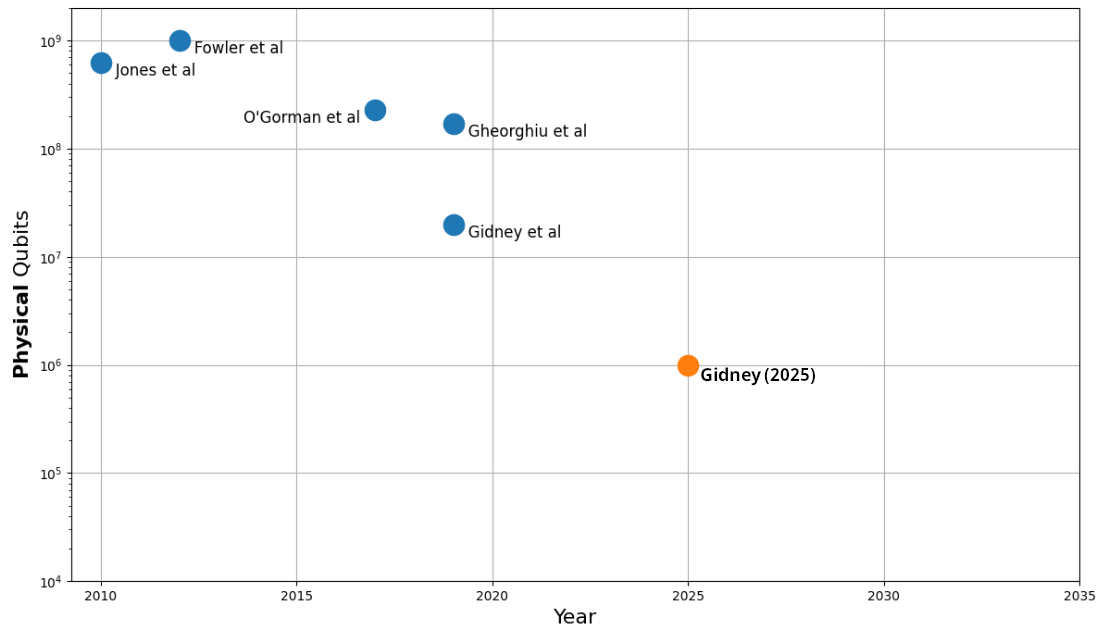}
	\caption{Historical estimates with comparable physical assumptions of the physical qubit cost of factoring 2048 bit RSA integers~\cite{gidney2025}.}
	\label{fig:gidney}
\end{figure}

Our world today is highly reliant on secure encryption. However, the advances in the
field of quantum computing threaten to break all currently prevalent public key
cryptography~\cite{djb_pqc}, an essential building block for cryptographic systems
establishing secure connections, e.\,g., VPNs. A Quantum
Computer~(QC) with a sufficient amount of logical qubits can break the underlying problems of factorization or discrete log based public key cryptography in (quantum) polynomial time~\cite{shor1997}. We call a QC with
sufficient power a Cryptographically Relevant Quantum Computer~(CRQC). Estimates regarding when a CRQC
will exist depend on both, the number of logical qubits that are necessary to break
the scheme, and also when this number of qubits will be achieved in real world
QCs. Figure~\ref{fig:gidney} depicts the trend of those forecasts regarding the
amount of necessary physical qubits to factorize a 2048 bit RSA
integer~\cite{gidney2025}. A report by the global risk institute indicates a
similarly worrying trend with a significant probability of QCs being able to break
2048 bit RSA in 24 hours during the next decade~\cite{qc_report}.
\newline
In response to this development, the US National Institute of Standards and
Technology~(NIST) designates most currently used public key algorithms
as deprecated or even disallowed after $2030\!-\!2035$\cite{nist8547}.
Instead, organizations should transition to algorithms of the field of post-quantum
cryptography~(PQC) which are expected to be secure also against strong quantum
computers.
European agencies, however, promote a similar schedule for introducing PQC with a
stronger focus on hybrid solutions that do not require as much trust in the long-term
security of the new PQC algorithms~\cite{eu_statement}. They also emphasize the
priority of protecting against store-now decrypt-later scenarios and devices with
long migration periods, such as public key infrastructures~(PKIs).
\newline
One common use case for PKIs is the authentication of VPN connections. There, we
encounter several specific challenges regarding the signature algorithms:
\begin{description}
	\item[Signature size:] The size should be small, so the network packet size
		increases as little as possible.
	\item[Amount of signatures per time:] As VPN gateways have to handle large volumes
		of VPN connections, creating many signatures quickly and efficiently must be
		possible.
	\item[Long term security:] The long migration period of those systems also means the
		new algorithms have to remain secure for long as migrating again is difficult.
\end{description}
Most of the algorithms standardized by NIST are module-lattice based which lack
confidence regarding their long-term security~\cite{bsi_kyber,ikev2_pqc_auth}. In
contrast, the family of hash-based signatures~(HBS) are solely based on the pre-image resistance of
the used hash-functions. This makes them ideal for long-term security~\cite{darzi2023}.
However, stateless HBS, e.\,g., SLH-DSA which is standardized by NIST, have large and
slow signatures making them unsuitable for VPNs. The stateful HBS~(S-HBS), on the
other hand, require state-management, as using the same part of the secret key twice
renders the whole key pair insecure. This seems impractical for the large quantities
of signatures in the VPN use case~\cite{nist800-208}.
\newline
To address the lack of long-term secure VPN PKIs, we propose a design consisting of
an S-HBS Certificate Authority~(CA) per VPN device. The CA signs classical
short-validity-time certificates within a predefined time schedule. Thus, the
large-quantity signatures use the cheaper classical algorithms, while the
S\mbox{-}HBS CA provides the long-term security anchor. The small signature size of
S-HBS further benefits our design. We selected four hours as an appropriate lifetime
for the classical certificates. Should QCs get closer to break our classical
algorithms that quick, we can still react and shorten their lifetime. Even switching
to fully quantum-secure leaf certificates later on is feasible, as the CA's strong
trust anchor allows for high levels of cryptographic agility in the leaf
certificates.
\newline
The predefined time-schedule means the state-management is static, as long as the
time on the machine remains correct. Thus the challenge is to make the design
resilient against faults and tampering regarding time management. For this, we
introduced several mechanisms, such as redundant timers or NTP servers, and
combined those with classical counter-based S-HBS management. We validated the
concept by implementing it within OpenIKED, an IPsec key exchange implementation
of the OpenBSD project. Our test results show that the quantum-resistant
implementation is resilient against a variety of state-tampering attempts while
introducing only limited performance costs.
\ifanonymous\else
\\~\\
\fi

\section{Related Work}
Several authors have addressed different approaches for secure state-management of
S-HBS, including time-based approaches, and also the integration of PQC into IPsec.
We now explore the respective state of the art and lay the groundwork for
understanding our design.
\newline
Wiggers et al.\ discuss in their current Internet Draft various aspects and security considerations of
state and backup management for S-HBS~\cite{wiggers}. They took a general approach and discussed
various different techniques, including time-based management, which
we focus on. They propose a system in which available signatures are
reserved for certain time windows, more specifically eight windows containing 128
signatures for a total of 1024 available signatures.
They identified several challenges regarding its time-based state-management
approach:
\begin {itemize}
	\item Accurate time keeping on the signing devices is essential. This is a serious challenge rooted in computer engineering.
	\item The index/state of the S-HBS can only increase, stepping back could cause a second signing with the same OTS key, which must not happen.
	\item The number of created signatures in each window has to be tracked to guarantee that no more key pair than allowed is used.
	\item The signing device has to maintain a proper state-management in general,
		i.e., update the state accordingly etc.
	\item The time keeping on the signing device has to be secure against malicious manipulation.
	\item In case of a replacement of the signing device the new device has to be in sync with the initial time keeping.
	\item An additional rate limit of the signatures as well as a second timer/clock could be useful.
\end{itemize}
As a backup Wiggers et al.\ suggest a record of the private keys, which includes the state, as well as a record of the windows and their configuration.
The disadvantage is the possible downtime resulting from waiting for the next time window, which could be bypassed by actively moving the time window.
This, however, forces the system to take the so-called clock drift into account for future signatures.
\newline
Regarding integrating PQC into IPsec or more specifically its key exchange protocol IKEv2, there exists a first formal analysis by Gazdag et al.~\cite{IKEv2-Anal}. It
analyses several RFC drafts and proves the effectiveness of the draft of
RFC~9242~\cite{interm_exch} which adds the new $IKE\_INTERMEDIATE$ exchange to IKEv2.
Herzinger et al.~\cite{RW-IPsec} evaluated the ideas of RFC~9242~\cite{interm_exch}
including additions for large KEMs like McEliece in a more practical setting. They
implemented the ideas in OpenBSD's IKED and highlighted the costs in terms of
performance and complexity. Bae et al.~\cite{ipsec_kems} also practically evaluated
the ideas and had similar results regarding the complexity. Yet, since they did not
introduce large KEMs, their performance results were significantly less worrying.
\newline
All those papers focused on ways to implement PQC key exchange mechanisms into the
IPsec stack. While this addresses the urgent topic of store-now decrypt-later
attacks, the problem of migrating PKIs for authentication remains mostly an open
research topic. It is currently rather touched on in standardization, in the form of
a RFC draft~\cite{ikev2_pqc_auth}. It introduces the PQC ML-DSA and SLH-DSA algorithms standardized by
NIST in \cite{fips204} and \cite{fips205}, respectively, to IKEv2: It highlights the overhead of using ML-DSA and even more of SLH-DSA
but also mentions the high confidence into SLH-DSA's security. S-HBS are, however,
not considered for IPsec authentication. We assume the reason to be the challenge of
managing their state.
\newline
A work specifically addressing S-HBS in the context of authentication is ``Post-Quantum Authentication in OpenSSL with Hash-Based Signatures'' by Butin et al.~\cite{xmss_openssl}.
The authors integrated XMSS, a standardized S-HBS algorithm~\cite{xmss}, into OpenSSL
and successfully used it for authentication.
They proved the feasibility of integrating S-HBS into existing infrastructure.
They did, however, conclude that the state-management can cause issues regarding TLS
integration and also increases its difficulty. Additionally to TLS, the authors also
tested S-HBS on S/MIME\@. They found S/MIME to be an easier use case of S-HBS, as it
requires less architectural changes.
Marzougui et al.~\cite{XMSS_Chain_Trust} even integrated XMSS together with Dilithium
into a chain of trust.
For that purpose they used and extended wolfSSL to perform a quantum-secure certificate-based TSL handshake.
In their chain of trust the CA was signed using Dilithium while the underlying certificates were XMSS certificates.
This approach differs substantially from ours, especially regarding the algorithm
choices and how we try to leverage the distinct features of S-HBS for the specific
context.

\section{Design}
The goal of our design is to leverage S-HBS' long term security and comparatively
small combined size of signature and public key for VPN authentication, while
mitigating the negative effects due to those algorithms being stateful. By selecting
S-HBS we inherently address the requirements \emph{small signature size} and
\emph{long term security}. However, the design must attend to the issues regarding
performance and statefulness to also meet the requirement of efficiently
creating a \emph{large amount of signatures per time} in order to sign many
time-critical VPN connection requests.

\subsection{General Concept and Idea}
\label{s:general_concept}
To guarantee a quantum-safe certificate-based authentication, we need to ensure a
quantum-safe chain of trust. The chain of trust is quite shallow, with one CA and one
layer of underlying certificates per VPN device.
The CA contains one XMSS key pair and signs itself with the first available signature
of its private key, which corresponds to key index 0.
As Subject Public Key (SPK) the CA contains the XMSS-root used to authenticate this
PKI's leaf certificates. Thus, the anchor of our chain of trust is based on XMSS whose
high level of (post-quantum) security makes it well suited for that use case.
\newline
\begin{figure}[t]
\centering
\begin{tcolorbox}[colback=gray!10, colframe=black!60, boxrule=0.6pt, width=\linewidth]
\begingroup
\renewcommand{\arraystretch}{1.4}
\begin{tabularx}{\linewidth}{@{} l X r @{}}
	\textbf{Certificate Field} & \textbf{Value} & \textbf{Size (Bytes)} \\
  \midrule
  Version & 3 & 3\\
  Serial Number & 00 & 2\\
  Signature Algorithm & ecdsaWithSHA256 & 10\\
  Issuer & CN=ExampleCA & 18\\
  Validity & Not Before: 2025-08-12 16:14:35 & 15\\
           & Not After:  2025-08-12 20:14:35 & 15\\
  Subject & CN=ECDSACrt & 20\\
  Subject Public Key Info: & Algorithm: Elliptic Curve & 7\\
                          & Key Parameters: secp256r1 & 8\\
                           & Public Key: [BIT STRING] & 66\\
  Extensions & X509v3 Subject Alternative Name: & 24\\
             & \quad DNS: example.com & 13\\
  Algorithm ID & 1.3.6.1.5.5.7.6.34 & 12\\
  Signature & [BIT STRING] & 2693\\
\end{tabularx}
\endgroup
\end{tcolorbox}
\caption{Basic structure of the XMSS-signed leaf certificates generated with
\texttt{dumpasn1}.}
\label{fig:rsa_cert_struct}
\end{figure}%
Unlike the CA, the leaf certificates are ordinary ECDSA certificates, as depicted in
Figure~\ref{fig:rsa_cert_struct}.
The CA's XMSS tree uses its remaining leaves to sign those certificates.
This ensures non-forgeability of leaf certificates, even against a CRQC\@.
The underlying certificates then use their ECDSA private key to sign and thus
authenticate the VPN connection handshakes.
Signing connections using ECDSA is theoretically not quantum-safe, as an attacker
with a CRQC could calculate the private key and, subsequently,
impersonate the VPN participant. However, the attack must be successful \emph{before}
the validity time of the leaf certificates has expired. Afterwards their private key
is useless. This has two implications:
\begin{itemize}
	\item The attacker must have the CRQC available for the attack.
	\item The QC must be able to break the classical algorithm quicker than the given
		validity period.
\end{itemize}
There are several estimates as to how long it would take a CRQC to break common
schemes like ECDSA or RSA\@. Yet, even optimistic estimates taking a QC with
1,000,000 physical qubits into consideration, assume the time to break RSA to be slightly under a week~\cite{gidney2025}, ECDSA being within a similar time span. Thus, with a
lifetime of, in our case, four hours, we expect our design to be secure for probably
decades. Also we gain high amounts of crypto-agility for the leaf certificates, as
they are swapped every four hours anyway. This means adapting both their lifetime or
their algorithms requires only the VPN software to support the new parameters; all
other trust comes from the still valid XMSS CA.
\newline
One single XMSS private key can sign $2^h\mkern-2mu -\mkern-2mu 1$ certificates, where $h$ is the height of the tree. We opted for the common tree
height $h = 16$. This gives us a total of $2^{16}\mkern-2mu -\mkern-2mu 1 = 65\, 535$
possible XMSS-signed certificates.
In theory and with a certificate validity of four hours, this leads to a CA lifetime
of $65\, 535 \cdot 4h \approx 29.925a$.
\newline
All involved certificates use the X.509 standard and Abstract Syntax Notation One (ASN.1).
Figure~\ref{fig:chain_of_trust} shows the general chain of trust, while
Figure~\ref{fig:chain_of_trust_xmss} depicts the certificates from the point of the
XMSS-tree.
It is also important to note that usually only one valid ECDSA certificate exists at
a time. So the underlying certificates depicted in the figures are separated chronologically.
\newline
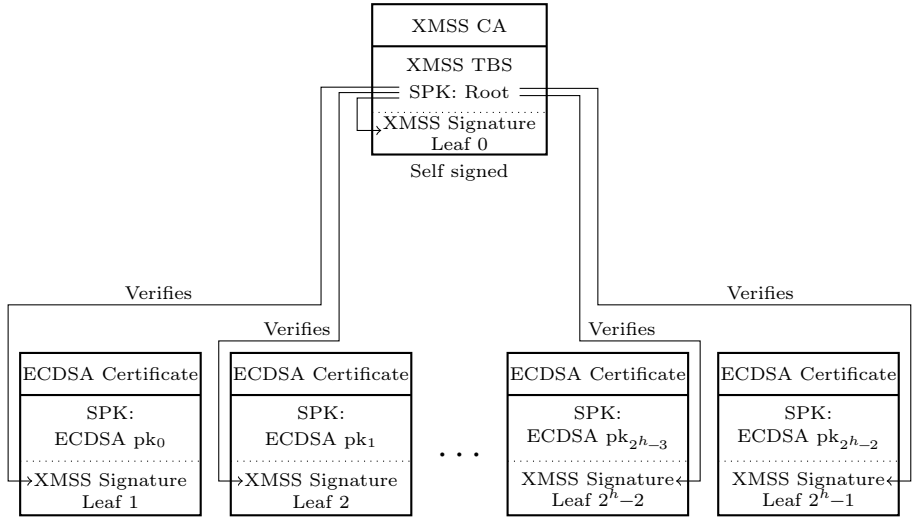
\begin{figure}[t]
\centering
\begin{tikzpicture}[scale=0.8]
	\draw[thick] (-1.45, 2) rectangle (1.45, -0.5);
	\draw[thick] (-1.45, 1.3) -- (1.45, 1.3);
	\draw[dotted] (-1.45, 0.2) -- (1.45, 0.2);
	\node[draw=none] at (0, 1.65) {\scriptsize XMSS CA};
	\node[draw=none] at (0, 1.0) {\scriptsize XMSS TBS};
	\node[draw=none] at (0, 0.55) {\scriptsize SPK: Root};
	\node[draw=none, align=center, font=\scriptsize] at (0, -0.15) {\scriptsize XMSS Signature\\Leaf 0};

	\draw[thick] (-7.3, -3.8) rectangle (-4.3, -6.5);
	\draw[thick] (-7.3, -4.5) -- (-4.3, -4.5);
	\draw[dotted] (-7.3, -5.6) -- (-4.3, -5.6);
	\node[draw=none] at (-5.8, -4.15) {\scriptsize ECDSA Certificate};
	\node[draw=none] at (-5.8, -4.8) {\scriptsize SPK:};
	\node[draw=none] at (-5.8, -5.25) {\scriptsize ECDSA pk$_0$};
	\node[draw=none, align=center, font=\scriptsize] at (-5.8, -5.95) {\\[0.4em]XMSS Signature\\Leaf 1};

	\draw[thick] (-3.8, -3.8) rectangle (-0.8, -6.5);
	\draw[thick] (-3.8, -4.5) -- (-0.8, -4.5);
	\draw[dotted] (-3.8, -5.6) -- (-0.8, -5.6);
	\node[draw=none] at (-2.3, -4.15) {\scriptsize ECDSA Certificate};
	\node[draw=none] at (-2.3, -4.8) {\scriptsize SPK:};
	\node[draw=none] at (-2.3, -5.25) {\scriptsize ECDSA pk$_1$};
	\node[draw=none, align=center, font=\scriptsize] at (-2.3, -5.95) {\\[0.4em]XMSS Signature\\Leaf 2};

	\draw[thick] (0.8, -3.8) rectangle (3.8, -6.5);
	\draw[thick] (0.8, -4.5) -- (3.8, -4.5);
	\draw[dotted] (0.8, -5.6) -- (3.8, -5.6);
	\node[draw=none] at (2.3, -4.15) {\scriptsize ECDSA Certificate};
	\node[draw=none] at (2.3, -4.8) {\scriptsize SPK:};
	\node[draw=none] at (2.3, -5.25) {\scriptsize ECDSA pk$_{2^{h}\mkern-3mu - \mkern-2mu 3}$};
	\node[draw=none, align=center, font=\scriptsize] at (2.3, -5.95) {\\[0.4em]XMSS Signature \\ Leaf $2^{h}\mkern-6mu - \mkern-5mu 2$};

	\draw[thick] (4.3, -3.8) rectangle (7.3, -6.5);
	\draw[thick] (4.3, -4.5) -- (7.3, -4.5);
	\draw[dotted] (4.3, -5.6) -- (7.3, -5.6);
	\node[draw=none] at (5.8, -4.15) {\scriptsize ECDSA Certificate};
	\node[draw=none] at (5.8, -4.8) {\scriptsize SPK:};
	\node[draw=none] at (5.8, -5.25) {\scriptsize ECDSA pk$_{2^{h}\mkern-3mu - \mkern-2mu 2}$};
	\node[draw=none, align=center, font=\scriptsize] at (5.8, -5.95) {\\[0.4em]XMSS Signature \\ Leaf $2^{h}\mkern-6mu - \mkern-5mu 1$};

	\draw[->] (-1, 0.44) -- (-1.7, 0.44) -- (-1.7, -0.1) -- (-1.3, -0.1);
	\draw[->] (-1, 0.62) -- (-2.3, 0.62) -- (-2.3, -3) -- (-7.5, -3) -- (-7.5, -5.93) -- (-7.1, -5.93);
	\draw[->] (-1, 0.53) -- (-2, 0.53) -- (-2, -3.6) -- (-4, -3.6) -- (-4, -5.93) -- (-3.6, -5.93);
	\draw[->] (1, 0.48) -- (2, 0.48) -- (2, -3.6) -- (4, -3.6) -- (4, -5.93) -- (3.6, -5.93);
	\draw[->] (1, 0.6) -- (2.3, 0.6) -- (2.3, -3) -- (7.5, -3) -- (7.5, -5.93) -- (7.1, -5.93);
	
	\node[draw=none] at (0, -0.8) {\scriptsize Self signed};
	\node[draw=none] at (-5, -2.8) {\scriptsize Verifies};
	\node[draw=none] at (-2.7, -3.4) {\scriptsize Verifies};
	\node[draw=none] at (2.7, -3.4) {\scriptsize Verifies};
	\node[draw=none] at (5, -2.8) {\scriptsize Verifies};
	\node[draw=none, font=\Large] at (0, -5.5) {$\cdots$};
	\end{tikzpicture}
	\caption{The chain of trust depicted as the PKI-tree.}
	\label{fig:chain_of_trust}
\end{figure}%
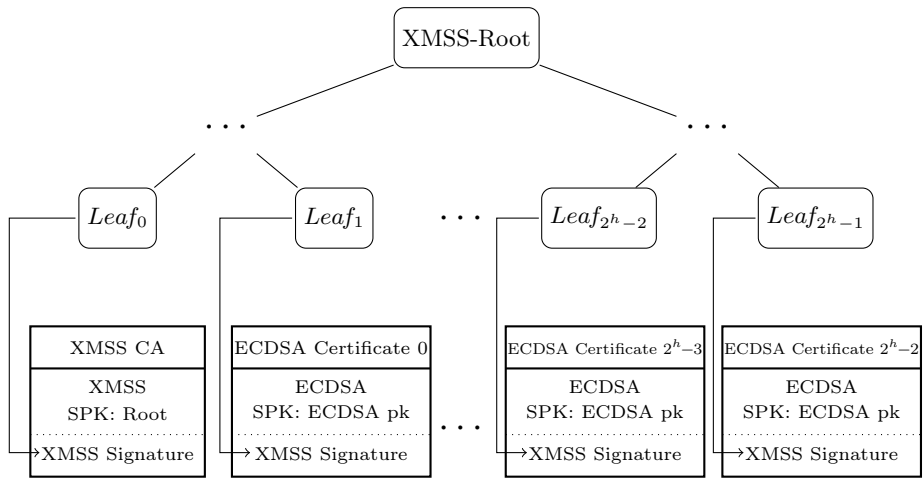
\begin{figure}[t]
\centering
\begin{tikzpicture}[
    scale=0.8,
    every node/.style={draw, rectangle, rounded corners, minimum size=0.8cm, text centered},
    level distance=1.5cm,
    level 1/.style={sibling distance=8cm},
    level 2/.style={sibling distance=3.6cm},
    level 3/.style={sibling distance=1.8cm}
]
	\node {XMSS-Root}
	    child {node[draw=none, font=\Large] {$\cdots$}
		child {node {$Lea\mkern-2mu f_0$}}
		child {node {$Lea\mkern-2mu f_1$}}
	    }
	    child {node[draw=none, font=\Large] {$\cdots$}
		child {node {$Lea\mkern-2mu f_{2^{h}-2}$}}
		child {node {$Lea\mkern-2mu f_{2^{h}-1}$}}
	    };
	\node[draw=none, font=\Large] at (-0.1, -3) {$\cdots$};
	\draw[thick] (-7.25, -4.8) rectangle (-4.35, -7.3);
	\draw[thick] (-7.25, -5.5) -- (-4.35, -5.5);
	\draw[dotted] (-7.25, -6.6) -- (-4.35, -6.6);
	\node[draw=none] at (-5.8, -5.15) {\scriptsize XMSS CA};
	\node[draw=none] at (-5.8, -5.8) {\scriptsize XMSS};
	\node[draw=none] at (-5.8, -6.25) {\scriptsize SPK: Root};
	\node[draw=none] at (-5.8, -6.95) {\scriptsize XMSS Signature};
	\draw[thick] (-3.9, -4.8) rectangle (-0.6, -7.3);
	\draw[thick] (-3.9, -5.5) -- (-0.6, -5.5);
	\draw[dotted] (-3.9, -6.6) -- (-0.6, -6.6);
	\node[draw=none] at (-2.25, -5.15) {\scriptsize ECDSA Certificate $0$};
	\node[draw=none] at (-2.25, -5.8) {\scriptsize ECDSA};
	\node[draw=none] at (-2.25, -6.25) {\scriptsize SPK: ECDSA pk};
	\node[draw=none] at (-2.25, -6.95) {\scriptsize XMSS Signature};
	\draw[thick] (0.65, -4.8) rectangle (3.95, -7.3);
	\draw[thick] (0.65, -5.5) -- (3.95, -5.5);
	\draw[dotted] (0.65, -6.6) -- (3.95, -6.6);
	\node[draw=none, scale = 0.85] at (2.3, -5.15) {\scriptsize ECDSA Certificate $2^{h}\mkern-6mu - \mkern-5mu 3$};
	\node[draw=none] at (2.3, -5.8) {\scriptsize ECDSA};
	\node[draw=none] at (2.3, -6.25) {\scriptsize SPK: ECDSA pk};
	\node[draw=none] at (2.3, -6.95) {\scriptsize XMSS Signature};
	\draw[thick] (4.25, -4.8) rectangle (7.55, -7.3);
	\draw[thick] (4.25, -5.5) -- (7.55, -5.5);
	\draw[dotted] (4.25, -6.6) -- (7.55, -6.6);
	\node[draw=none,scale = 0.85] at (5.9, -5.15) {\scriptsize ECDSA Certificate $2^{h}\mkern-6mu - \mkern-5mu 2$};
	\node[draw=none] at (5.9, -5.8) {\scriptsize ECDSA};
	\node[draw=none] at (5.9, -6.25) {\scriptsize SPK: ECDSA pk};
	\node[draw=none] at (5.9, -6.95) {\scriptsize XMSS Signature};

	\node[draw=none, font=\Large] at (-0.1, -6.5) {$\cdots$};

	\draw[->] (-6.5,-3) -- (-7.6,-3) -- (-7.6,-6.9) -- (-7.1, -6.9);
	\draw[->] (-2.9, -3) -- (-4.1,-3) -- (-4.1, -6.9) -- (-3.6, -6.9);
	\draw[->] (1.2, -3) -- (0.5, -3) -- (0.5, -6.9) -- (1, -6.9);
	\draw[->] (4.8, -3) -- (4.1, -3) -- (4.1, -6.9) -- (4.6, -6.9);
\end{tikzpicture}
\caption{The chain of trust from the XMSS-tree's perspective.}
\label{fig:chain_of_trust_xmss}
\end{figure}%
One aspect to consider is that if the interval of creating new certificates is
identical to the validity time of the certificates, it could lead to certificates not
being created in time.
To prevent this, we set the certificate creation time interval slightly
shorter (we chose a difference of two minutes) than the certificates' validity time.
Hence, for a short period of time we have two valid certificates. We expect this to
not significantly impact any other considerations.
\newline
In theory, the scheme with our parameters lasts for almost 30 years. In practice, we
expect that within such a time frame there always occur irregularities regarding time
management. We implement measures to prevent those from weakening the security at the
cost of sometimes skipping parts of the private key. Thus, the real lifetime of one
CA will be shorter. Yet, we assume that it will easily suffice for the common CA
lifetimes of ten years. Hence, the longevity of the scheme is adequate and also has
margin to support a shorter validity time for the leaf certificates. If necessary,
the design would also allow for optimization for smaller signatures but decreasing
the life time and vice versa, by adjusting the tree height $h$.
\newline
In summary, we designed an architecture for post-quantum authentication of a VPN
device. The device uses an XMSS CA as its long term security anchor. This CA issues
short-time ECDSA certificates which then authenticate the VPN connections. The
post-quantum security of the certificates bases on the assumption that all the first
generations of CRQCs will not be able to break the cryptography within the short
time frame. Additionally, we can always just swap the leaf certificate algorithm.
Finally, the state-management benefits from the fixed schedule which predetermines
exactly which XMSS index will issue a certificate for which time interval. Thus, with
a correct system time, the state-management is entirely static.

\subsection{Robust Time-Based state-management}
The time-based approach greatly simplifies the S-HBS state-management but the time
dependency also introduces a new attack vector when compared to stateless signature
schemes. Namely, if an attacker manages to manipulate the system time, the security
of the design is at risk. More specifically, there are two effects an attacker could
try to achieve:
\begin{description}
	\item[Double Signing:] In this case, the attacker manages to change the device
		time so it uses an XMSS index which was already used in the past. As they
		already have knowledge about the used secret key information, they can use this
		index to forge signatures. Thus, this must be prevented. We rely on the XMSS
		implementation preventing the index from decreasing and, hence, from double
		signing.
	\item[Key exhaustion:] Instead of trying to decrease the index, an attacker could
		also try to increase it to the maximum, thereby exhausting the key. This leads to
		a Denial of Service~(DoS) attack.
		Wiggers et al.~\cite{wiggers} also mention this risk and recommend adding a rate
		limiter to the signatures as well as a second clock.
\end{description}
The more dangerous effect of double signing is already taken care of by the XMSS
implementation itself. Still, an attacker can use multiple time manipulation
techniques to mount DoS attacks. This means that the design needs to be resilient
against state-tampering but also have strategies that make it robust in the case of
successful system time manipulation for it to provide a dependable VPN service. We
introduced several different approaches that increase the designs resilience.

\subsubsection{Signed Time-Schedule}
When generating the XMSS key pair, we also create a schedule which exactly defines
the time of validity for each XMSS index. By that we create a strong dependency
between the time-state, represented by the current timestamp, and the XMSS-state,
represented by the XMSS index. Before any signing operation, the system checks the
schedule for the current time interval and if the current XMSS index matches the one
in the schedule. Hereby, it
recognizes if one of the two states deviates from the planned schedule.
\begin{figure}[t]
\centering
\begin{tabular}{|c|c|c|c|} \cline{1-4}
\multicolumn{1}{|c|}{\textbf{8 Bytes}} & \multicolumn{1}{c|}{\textbf{2 Bytes}} & \multicolumn{1}{c|}{\textbf{4 Bytes}} & \multicolumn{1}{c|}{\textbf{4 Bytes}} \\ \cline{1-4}
\textbf{Creation Date} & \textbf{Validity Time} & \textbf{Current Index} & \textbf{Max Index} \\
\hline
1751720001 (time\_t) & 240 (int16\_t) & 2 (int) & 65536 (int) \\
\hline
\multicolumn{4}{|c|}{\textbf{Signature (2692 Bytes)}} \\
\hline
\end{tabular}
\vspace{0.5em}
\caption{Structure of the schedule with example values. The date is a POSIX timestamp
and the validity time is given in minutes.\label{fig:schedule}}
\end{figure}%
\newline
A primitive solution for the schedule would be a timetable with a time delta value
for each index. However with the amount of indices, the table would become excessively
large. Instead, we only store its creation time, the time interval, the index of the
first certificate, and the maximum index. Figure~\ref{fig:schedule} shows the
resulting structure. This is sufficient to calculate the exact
timestamps. To protect it from any tampering, we sign the schedule with the first
available XMSS index $i$. For the initial setup, this means $i = 0$ signs the CA
itself, $i = 1$ signs the schedule, and $i = 2$ signs the first leaf certificate.
This results in the \texttt{starting index} field of the schedule being two initially.
We also distribute this schedule to our VPN peers using an out-of-band channel.
As the distribution of the file is similar to other auxiliary information for
certificates, e.\,g., Certificate Revocation Lists~(CRLs), it should be able to
utilize similar mechanisms.
\newline
The strong dependency between the two kinds of state allows to detect any anomalies
regarding one of the two states. This could be that the time changed, e.\,g., due to
manipulation, or that the index is at the wrong place, for example because of a
system downtime or crash. It also prevents the device from repeatedly creating
certificates in case of a malfunction in the relative timekeeping which would lead to
key exhaustion. Instead, should the time suddenly change in a way that the XMSS index
does not fit the schedule anymore, the system recognizes the mismatch between the two
states. Our design then proposes that the system should not create a certificate but
instead notify an administrator to manually verify both states and correct any wrong
ones.
\newline
An administrator correcting the time is straightforward. Correcting the index needs
attention within our design. There we have two possibilities. (1)~The XMSS index
being stale and (2)~the index being ahead of schedule. The second case can only
happen if an attacker manages to set the system time to the future, without the
device noticing or with the administrator accepting this time jump. In that case, we
face two challenges:
\ifanonymous\else
\newpage
\fi
\begin{itemize}
  \item The XMSS private key for the scheduled indices might already be used.
		Until we are in schedule again, we probably cannot provide the VPN service.
	\item A drift to a timestamp that actually lies in the future causes the VPN device
		to create certificates that are not yet valid but will be in the future. An
		attacker could see the public key information, use a CRQC to break it until the
		certificate is valid, and then impersonate the VPN device.
\end{itemize}
To address those issues, we propose a mechanism to update the schedule. For the
update, we just write a new schedule starting from the current index $i + 1$ and the
current timestamp. The device signs the table using index $i$ and, again, distributes
it to its peers via the out-of-band channel. Then, the device can use index $i + 1$
to sign the next leaf certificate and continue its work from there. This both
fixes the first, and also the second concern, as long as all peers receive the
current schedule. With that they can, before verifying a certificate, first make
sure that its index matches the time interval in the schedule. Thus, they will
reject the prematurely generated certificates.
\newline
If we are behind schedule we can easily address it by creating dummy signatures to
increase the index. However, for the index being far behind schedule signing many
dummies wastes private key space. Instead, we could also use the schedule update.
There, we need to only invest one signature for signing the new schedule itself and
the overhead to distribute it to the peers. Another use case for the update mechanism
is the modification of the time interval. Should, e.\,g., a CRQC be on the horizon that
might manage to break the leaf certificates in the current time interval, we can just
change it by updating the schedule.

\subsubsection{NTP Server}
Instead of mitigating the effect of successful time tampering, like the schedule, we
can also try to harden the design and prevent time tampering in the first place.
One strategy for this is the addition of the Network Time Protocol (NTP).
NTP servers can help to reduce the dependency on the system time by fetching it from
designated NTP servers.
To use it, we need at least one NTP server but more are possible to increase reliability at the cost of complexity.
\newline
To verify that the system time is accurate and more resilient against manipulation,
the system fetches the accurate time from the NTP servers and updates its time
accordingly.
Yet, this also adds dependencies to the NTP server as a successful attack on the
server or the NTP connection again affects our state-management.
Thus, we propose to use more than one, ideally three NTP servers, similar to the principle of triple modular redundancy used in computing~\cite{tmr}.
There, we update our system time to the time of the NTP server with the smallest
delay, given that the majority of the servers' timestamps match.
If a more significant update is performed on the system time, the administrator gets
informed about the situation.
\newline
We also need to take a man-in-the-middle attack on the NTP connections into account.
Native NTP does not authenticate its connections, which makes such attacks feasible.
As mitigation we propose the usage of Network Time Security~(NTS). This uses TLS to
secure the connection. To achieve full post-quantum security we also need to consider
TLS\@. Similarly, the DNS name resolution must be post-quantum authenticated.
Both topics, however, are their own research track and will not be considered further
within this work.
Our implementation utilizes OpenBSD's native OpenNTP client, which does not support
NTS\@. So for a production-ready deployment, we need to switch to other software
projects, like \emph{chrony}\footnote{https://chrony-project.org}, or use dedicated
internal NTP servers or VPN-like protection of the NTP connection.

\subsubsection{Relative \& Absolute Time Keeping}
The time used to schedule the generation of new certificates can be measured in two different ways:
\begin{description}
	\item[Relative Timing:] means we run a timer for a given interval, e.\,g., four
		hours. After that interval and independently of the system time, it issues the
		creation of a new certificate.
	\item[Absolute Timing:] means the system generates a certificate as soon as its
		time reaches a certain time value.
\end{description}
We use a combination of the two approaches. That means after signing a new
certificate we set a relative timer of ca.\ four hours. At the same time we
calculate the absolute timestamp of the next certificate creation. After the relative
timer triggers the next certificate creation, the system compares this time stamp
with the current time. If they match a new certificate is being issued. If they do
not match, we compare with the schedule and issue a warning to the administrator to
make them verify the current state. If the system reboots, we lose the state of the
relative timer. So, in that case it does not offer additional protection.

\subsubsection{Conclusion of our Approaches}
In order to harden our design against time-tampering attacks, we introduced several
measures. We introduced time verification using a triple-redundant set of
NTP servers. As long as they use NTS to protect the NTP connections, the system time
becomes more resistant to tampering or faults. Otherwise it is for operations
to find ways to make the system time as stable as possible. The signed time schedule
and the introduction of separate timers allow to detect time anomalies and react to
them. Thus, the worst case of a successful time tampering attack is that the system
recognizes the anomaly and waits for administrative correction. Classical VPN setups
with certificate-based authentication suffer to a similar extent from those attacks,
as certificate validation also requires a correct system time.

\subsection{Security Analysis}
We will analyze the security and robustness of our design regarding time anomalies in
two separate scenarios: During operations and after a system downtime. In both cases
the system time can be either set to the future or the past.
\subsubsection{Scenario 1: Time Manipulation during Operations}
\label{s:scen_ops}
The first scenario is that the system clock gets set back or forth in time. After the
current relative time interval has passed and the next certificate creation is due,
the system compares its time with the estimated time. This mismatch and also the fact
that the XMSS index is two days ahead / behind of the schedule and
regarding the system time lead to the certificate issuing process halting. Then, the
administrators are informed. After they fix the system time, operations can continue as
usual.

\subsubsection{Scenario 2: Downtime}
The second scenario represents a reboot of the system with potentially an unknown
amount of downtime. In the case of this event, we face the following situation
regarding the states:
\begin{itemize}
	\item A reboot deletes the state of the relative timer. Hence, its state is
		undefined and we cannot take it into account.
	\item The time may be correct but can also be stuck in the past. In that case the
		NTP queries gain importance. However, also there we have to take NTP drift into
		account~\cite{wiggers}, i.\,e., that NTP tries to adjust the time gradually and not
		abruptly. Hence, it may take some time for the system time to match the
		NTP server's. Furthermore, also in this situation malicious time tampering is a
		possibility we need to take precautions against.
	\item The device still has the last leaf certificate on its disk. This indicates in
		what time interval it stopped working and, consulting the schedule, where the
		index should be.
	\item The index should be at the value matching the last used certificate or one
		ahead. If it is ahead of the certificate, the system needs to spontaneously
		halt during certificate generation. As the implementation first changes the index
		variable and then crafts the signature, interrupting this process leads to the
		index being wasted.
\end{itemize}
As soon as the NTP process has finished and the system time is stable, the device
uses its index and certificate to obtain the expected time value and compares that
with the actual system time. If it all adds up, i.\,e., the system was just quickly
rebooting, it can continue working as usual. Otherwise, we proceed as in
Section~\ref{s:scen_ops} and notify the administrator. Yet in contrast to
Section~\ref{s:scen_ops}, we now have to consider the scenario where the index
actually is behind schedule. In that case, after the administrator confirms the
correctness of the time, we either sign dummy messages until the index catches up, or
we update the schedule so we do not waste as much private key space. For the exact
threshold value we propose to update if we needed to create more than ten dummy
signatures.
\newline
If an administrator faultily accepts a wrong time update after a downtime or the time
gets manipulated gradually so the system does not skip indices, there are two
possibilities: The first is that we jump back in time. This leads to the issuing of
certificates that are already invalid from the start, causing the VPN connections to
fail until the administrator fixes the time and updates the schedule, again. A jump
forward in time is more problematic. There, the device issues certificates that will
be valid in the future but are not yet. This means an attacker gains more time to
attack the certificates which severely affects our security against a CRQC-empowered
attacker. Also the VPN connections right now will fail due to the certificates
not yet being valid. These faults should trigger the administrator to address the problem
and fix time and update the schedule.
\newline
An especially severe case of the downtime scenario is a switch of the devices, also
mentioned by Wiggers et al.~\cite{wiggers}. For this to work, one needs a reliable
copy of the private key and its state, e.\,g., by extracting it from the disk or
transferring the Hardware Security Module~(HSM).

\section{Discussion}
We presented a design that leverages the long term security and comparatively small
size of signature and public key of XMSS for VPN authentication. It uses a time-based
approach to simplify and harden the state-management. We also propose measures to
increase the robustness of the design against time manipulation. In our theoretical
security analysis, we demonstrate how such a system deals with possible attack or
fault scenarios. We showed that it detects time manipulation and can notify an
administrator, prompting them to fix the states. Even if that mechanism fails and the
system adjusts to a wrong time, the worst case is a DoS of the VPN which can be
easily recovered by an administrator. In that case it resembles just the effects of
classical VPN certificate authentication. The state-protection mechanisms, in
general, work on all possible VPN devices. Yet, they favor systems with little
downtimes, such as gateways, and is less suited for, e.\,g., mobile clients.
\newline
We implemented the design in OpenBSD's IKED and evaluated it using a setup using both
official and local NTP servers. Tests with the implementation proved our
assumptions for the scenarios. There, we use OpenNTP as NTP client which does not
support NTS, leading to a potential weakness for man-in-the-middle attacks on the NTP
connections, allowing for an easier time tampering. However, switching to a client
supporting NTS fixes the issue, as long as there is a PQC-capable NTS implementation.
Another potential class of vulnerabilities we found are Virtual Machine~(VM) snapshots.
By saving and later reloading the current state of the virtual machine, it is
possible to reset the XMSS private key state to this old state, potentially causing
double signing. To prevent this we recommend not running any VPN devices in VMs, as
it generally poses risks regarding the later recovery of randomness seeds.
Another generally useful option is to put the key on an HSM, further hardening it
against state-tampering.

\subsection*{Potential Improvements}
We showed that our design works for authenticating one VPN device by assigning it an
XMSS CA. Yet, often PKI structures for VPN authentication are in place to simplify
key handling by not having to manage per device keys/certificates. This comes from
the possibility to trust one CA that then signs and validates many systems. Another
similar use case is the authentication of redundant VPN devices, i.\,e., two devices
that fulfill the same use but still work even if one is down and also share load.
\newline
We see two approaches to address this. First, we can move the XMSS CA to a separate
system which issues short time leaf certificates to a fixed number of VPN devices.
This means that the certificates do not increase and with the tree height of 16 the
CA still has a long lifetime. With an increase in tree height we can still
drastically increase the signature volume of one CA with little increase in signature
size. The downside of this approach is the additional complexity of managing the
separate system and also securing the communication channel between the VPN devices
and the CA. The other option is to just add another layer of CAs which again is
time-based, resulting in a classical PKI tree structure just with one layer more as
each device has an intermediate CA itself. This means a less complex design but adds
more signatures decreasing VPN tunnel creation performance. We expect that both
approaches are valid for different use cases. e.\,g., redundant devices sharing one CA
seems practical, whereas large distributed infrastructures will require separate CA
layers. However, this needs more testing and validation, as well as approaches to
secure the newly resulting communication paths.
\newline
We practically and theoretically tested our design by playing out several
scenarios, yet, formally verifying it would further validate its
feasibility.

\subsection*{Alternatives}
Usually people opt for the stateless signature schemes, also in the context for PQC,
just as with the respective RFC Draft~\cite{ikev2_pqc_auth}. S-HBS, on the other
hand, are seen more as a niche solution~\cite{xmss_openssl}. Applying it to the
use case of VPN authentication is a novel approach. We introduced measures to mitigate
the challenges of secure state-management which also comes with additional
complexity. Thus, we must attend the question of advantages over the NIST-standardized algorithms ML-DSA and SLH-DSA\@.
\newline
One big disadvantage of ML-DSA, of course, is its dependence on the hardness of
certain mathematical problems regarding structured lattices. Advances in cryptanalysis regarding
ML-DSA might significantly weaken or even nullify its security assumptions. So even
if ML-DSA in its largest parameter set has the same NIST security category 5
(equivalent to the security of AES-256~\cite{nist8547}) as the HBS primitives, they
are probably not comparable in terms of long-term
security~\cite{bsi_kyber,ikev2_pqc_auth}. Thus, for a reliable long-term design, we
need to use HBS\@.
Another advantage of XMSS's little need for complex computations is an easier
implementation on HSMs compared to ML-DSA, as shown in~\cite{xmss-hms}.
\newline
\begin{figure}[t]
\centering
\begin{tabular}{lccc}
\toprule
\textbf{Scheme} & \textbf{Signature} & \textbf{Public Key} & \textbf{NIST Category}\\
\midrule
ML-DSA-44       & 2,420 B   & 1,312 B		& 2 \\
ML-DSA-65       & 3,309 B   & 1,952 B		& 3 \\
ML-DSA-87       & 4,627 B   & 2,592 B		& 5 \\
\midrule
XMSS\_10		& 2,500 B  & 68 B     & 5 \\
\textbf{XMSS\_16}		& \textbf{2,692 B}  & \textbf{68 B}     & \textbf{5} \\
XMSS\_20		& 2,820 B  & 68 B     & 5 \\
\midrule
SPHINCS+128s		& 7,856 B		& 32 B	& 1 \\
SPHINCS+192s		& 16,224 B	& 48 B	& 3 \\
SPHINCS+256s		& 29,792 B	& 64 B	& 5 \\
\bottomrule
\end{tabular}
\caption{Comparison of public key and signature sizes in the \texttt{liboqs} implementation in relation to the respective NIST security category for ML-DSA, XMSS, and SPHINCS+
which is the primitive for SLH-DSA\@. All HBS implementations use SHA-256 as hash
function for the comparison.}
\label{fig:dil_benchmark}
\end{figure}%
The other big benefit of S-HBS like XMSS is its signature and public key size.
Especially the signature is part of the leaf certificate and, therefore, must be
transferred for each VPN key exchange. Figure~\ref{fig:dil_benchmark} shows that
already compared ML-DSA, XMSS is significantly smaller for a comparable security. A
comparison to SLH-DSA increases the advantage of XMSS to a factor of more than ten.

\section{Conclusion}
The recent advances in quantum computing demand urgent action to migrate systems to
post-quantum alternatives, including certificates for VPN authentication. While the
straightforward way is to use the NIST candidates as sketched
in~\cite{ikev2_pqc_auth}, this introduces potential long-term weaknesses and
drastically increases the amount of data to be sent for each VPN connection attempt,
thereby worsening VPN performance. S-HBS have less of those disadvantages but are often shied
away from due to their need for state-management and limited amount of signatures. To
address this issue, we proposed a design using an XMSS based CA issuing short-time
classical certificates. We considered the whole system quantum-safe as long as no CRQC
is able to break the classical algorithm quickly enough before it becomes invalid.
Another benefit of the design is the cryptographic agility of the leaf certificates,
as we can easily change their cryptographic parameters or even validity time. We
addressed the challenge of state-management by creating a fixed time-schedule for
each XMSS state, binding it to the time. Thus, the state-management becomes easier
but we had to harden the system against time-tampering and prepare strategies to deal
with a wrong system time, also after system downtime.
\newline
We implemented and tested the design based on OpenIKED, the Internet key exchange
daemon of OpenBSD\@. Additionally we drafted and implemented measures to harden it
against time tampering, including a signed schedule, triple-redundant
NTP servers~(still requiring a NTS-capable implementation), and a combination of
relative and absolute timers. As the measures profit from a little amount of reboots,
the design is especially viable for the authentication of VPN gateways.
In summary, we contribute:
\begin{itemize}
	\item A robust novel design for quantum-secure VPN authentication, based on a
		combination of XMSS and classical cryptography, mitigating XMSS state-management
		by statically binding it to the time.
	\item A scenario-based theoretical security analysis of the design.
	\item A proof-of-concept implementation based on OpenBSD's OpenIKED validating our
		findings.
\end{itemize}
For future work, we plan the design's adaptation for larger, distributed
infrastructures, including more complex PKI trees. Also, we aim to compare using one
time-based XMSS CA for several devices with adding more CA hierarchies. Furthermore,
we plan to formally validate the security properties of the resulting design.

%
%
%
%
\bibliographystyle{splncs04}
\bibliography{bib}

\end{document}